\documentclass[11pt,titlepage]{article}

\usepackage{jinstpub} 

\title{\boldmath  On the Use of Neural Networks for Energy Reconstruction in High-granularity Calorimeters}

\author{N.~Akchurin, C.~Cowden, J.~Damgov, A.~Hussain, and S.~Kunori }

\affiliation{Texas Tech University, Department of Physics and Astronomy,\\
Advanced Particle Detector Laboratory \\ Lubbock, TX, 79409, USA}

\abstract{
We contrasted the performance of deep neural networks - Convolutional Neural Network (CNN) and Graph Neural Network (GNN) - to current state of the art energy regression methods in a finely 3D-segmented calorimeter simulated by GEANT4.  This comparative benchmark gives us some insight to assess the particular latent signals neural network methods exploit to achieve superior resolution.  A CNN trained solely on a pure sample of pions achieved substantial improvement in the energy resolution for both single pions and jets over the conventional approaches.  It maintained good performance for electron and photon reconstruction.  We also used the Graph Neural Network (GNN) with edge convolution to assess the importance of timing information in the shower development for improved energy reconstruction.  We implement a simple simulation based correction to the energy sum derived from the fraction of energy deposited in the electromagnetic shower component.  This serves as an approximate dual-readout analogue for our benchmark comparison.  Although this study does not include the simulation of detector effects, such as electronic noise, the margin of improvement seems robust enough to suggest these benefits will endure in real-world application.  We also find reason to infer that the CNN/GNN methods leverage latent features that concur with our current understanding of the physics of calorimeter measurement.
}


\begin{document}
\maketitle
\flushbottom
%%%%%%%%%%%%%%%%%%%%%%%%%%%%%%%%%%%%
\newpage
\section{Introduction}
\label{sec:ML}
Calorimeters have played crucial roles in high energy physics experiments for decades.  In the last two, new ideas and techniques, such as dual-readout~\cite{AKCHURIN2005537} and particle flow~\cite{Sefkow:2015hna,Sirunyan_2017}, emerged and contributed to the increased understanding of calorimeter fundamentals.  Future accelerators, such as the high luminosity LHC or the FCC, push energy and luminosity bounds beyond our capacity today.  We must design calorimeters with increased granularity, fast timing, and high degrees of radiation tolerance to perform at future accelerators.  
%These, for example, include high granularity, fast timing, and radiation tolerance.

Those involved in calorimeter design have known for some time that devices often exhibit different characteristic responses to different particle types.  For instance, the different responses to electromagnetic and hadronic initiating particles, whose ratio is usually denoted as $e/h$, has led to many arguments to the merits of compensating ($e/h=1$) {\it vs} non-compensating ($e/h \neq 1$) calorimeter designs.
Since showers initiated by hadrons contain both electromagnetic (EM) and hadronic components, the fluctuation of the fraction of the initial energy into the EM component ($f_{\rm em}$) drives much of the observed variance of the energy measurement.
This insight, for instance, served as the impetus for the dual readout methodology in which a calorimeter with two readout types - with two different $e/h$ ratios - allows experimenters to determine $f_{\rm em}$ on an event-by-event basis.  One can measure energy much more precisely knowing $f_{\rm em}$ than not.
However, other less significant response fractions still exist; for example, the fraction of the initial energy going undetected in inelastic nuclear interactions.
Capturing these additional sources of fluctuations would require additional capabilities.

Interestingly, processes involved in the shower of particles inside a calorimeter have corresponding distributions of deposited energy.
Experimentalists have overlooked this fact since previous calorimeters lack the spatial granularity to warrant the development of reconstruction algorithms so finely attuned to local regions of showers.
The shear complexity of a reconstruction algorithm which could take advantage of such high granularity has also impeded much progress in this approach.

Recently the CMS experiment has adopted a high granularity calorimeter design as part of its end-cap upgrade ~\cite{Collaboration:2293646}. 
This has prompted further developments in reconstruction techniques which employ advanced pattern recognition techniques such as image processing.

The advance of the Convolutional Neural Network (CNN)~\cite{Fukushima1979,lecun2015deeplearning}, and neural networks in general, can allow us to more elegantly solve this challenging problem of reconstruction. 
We can take advantage of the CNN's sensitivity to spatially distributed signals to develop a model that accurately estimates the initiating particle's energy from a high granularity calorimeter.
Graph Neural Networks (GNN)~\cite{9046288} may potentially have advantages over the CNN since the GNN can accommodate a variety of geometries that a CNN cannot.  This detail implies that a GNN may be better suited to a full size experimental detector geometry.
In fact, many researchers have already studied applications of these techniques in the context of HEP collider experiments.
The application of CNNs has been tested in simulation for energy reconstruction on single pions~\cite{neubuser2021optimising, Belayneh_2020} as well more general detector settings~\cite{aleksa2019calorimeters}.
Several studies have investigated GNNs for the clustering as part of the particle flow application with multiple pions~\cite{Qasim_2019, Kieseler_2020, qasim2021multiparticle, ju2020graph} and jet tagging \cite{Qu_2020}. Another study investigated the separation between neutral and charged hadrons depositions within the same calorimeter volume~\cite{Di_Bello_2021}. We also note that neural networks have been successfully applied for particle identification~\cite{de_Oliveira_2020, Belayneh_2020}. 

%% new concluding paragraph
We studied in detail the performance of a CNN applied to energy regression in a finely 3D-segmented calorimeter simulation.
We also introduce a GNN (GNN is a popular choice for detectors with irregularly shaped geometry and sparse signal distribution in the calorimeter volume) into this study to evaluate the contribution of precise timing to energy reconstruction performance. 
The GNN, in this particular case, gives us convolutional functionality similar to the CNN with the addition of properly structured time information.
Although, CNNs can ingest timing data as a ``color'' dimension, this leads to unsatisfying results since time is not a nominal variable.
In our investigation of energy reconstruction using CNN/GNN techniques, we see promising improvements in resolution beyond what has been demonstrated by dual, and proposed triple, readout calorimeters~\cite{GROOM2007633}.
We intend to use these investigations as a stepping stone in the development of state-of-the-art calorimeters for future experiments.
Comparison of neural network architectures and their application to the particle flow algorithm remains outside of the scope of this study.    

In the next sections, we shall describe our simulation study and the reconstruction methods that we examined.  In Section~\ref{sec:perf}, we describe the resulting performance.
We describe the estimation of $f_{\rm em}$ with a CNN in Section~\ref{sec:CNNfem}, and we describe a further study showing the CNN's ability to leverage topological information to improve resolution in a compensating calorimeter in Section~\ref{sec:USi}.  In Section~\ref{sec:timing}, we introduce timing information and reconstruct energy with a GNN.
%% new concluding paragraph

\section{Simulation Study}
%\section{Calorimeter Setup and Methods for Energy Reconstruction}

%\subsection{Standalone Setup and Simulation}

We simulated
a sampling calorimeter with alternating plates of 17 mm copper (absorber) and 3 mm silicon (active material) with GEANT4~\cite{GEANT4:2002zbu} version 4.10.06 and we used the FTFP\_BERT physics list to describe the hadronic interactions. The calorimeter is 1.5 m deep (8.8 interaction lengths) and covers 1$\times$1 m$^2$ in transverse direction. The volume is divided in $2\times2\times2$ cm$^3$ cells, representing individual readout channels. The minimum ionizing particle (MIP) signal is about 1.0 MeV per cell and 
we set the minimum energy threshold at
0.6 MeV/cell in this study. Electronics noise is not taken into account.
The position of incident particles is smeared in a $4\times4$ cm$^2$ in transverse plane.

\section{Benchmark Energy Reconstruction Methodology}
%\subsection{Energy Reconstruction: Simple Sum, EM-correction, CNN Prediction}

We used two traditional energy reconstruction techniques to benchmark performance of current methods.
%We compared the CNN performance to with two traditional energy reconstruction techniques.
The first and simplest approach is the linear sum of all deposited energy where the active material is calibrated using electrons.
In the second technique, we employ to benchmark the potential performance of a comparable dual-readout calorimeter.
In it, we exploited information from the GEANT4 simulation to correct the energy sum for the $f_{\rm em}$ of the shower on an event-by-event basis.

\begin{figure}[ht]
\centering
\includegraphics[width=0.49\textwidth]{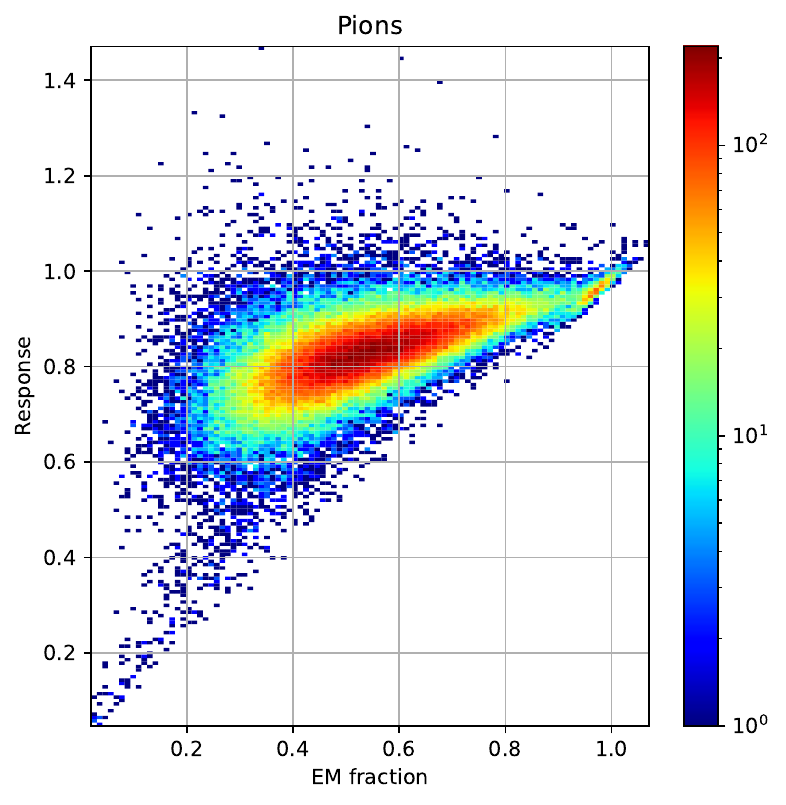}
\includegraphics[width=0.49\textwidth]{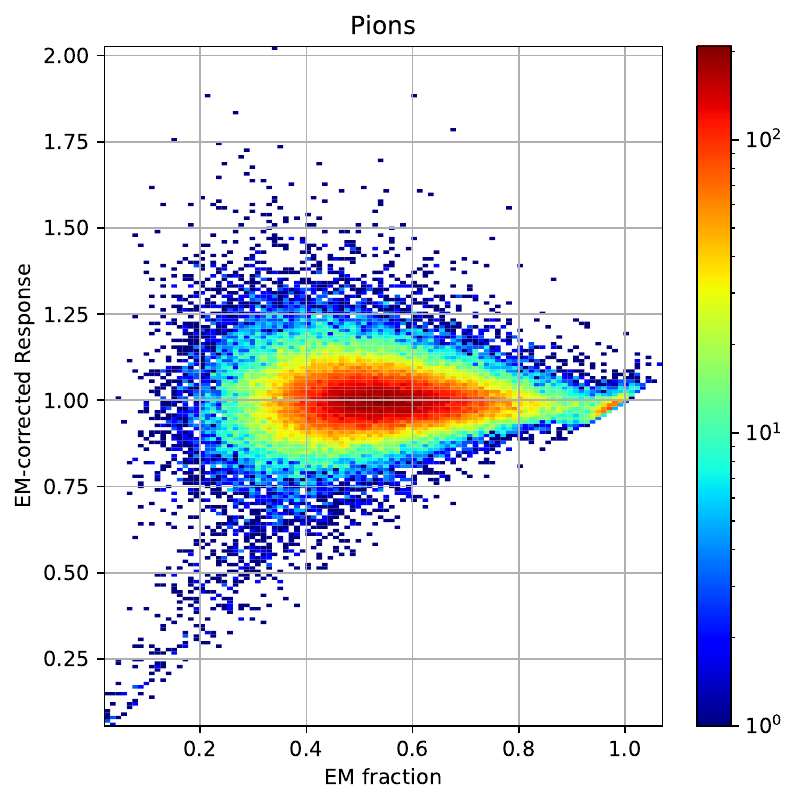}
\caption{The calorimeter response: simple energy sum over all channels (left) and $f_{\rm em}$ corrected energy sum (right) for pions with energy from 5 to 150 GeV. Entries with low EM and low response are from events with low showering activity and MIP-like tracks in the calorimeter volume.}
\label{fig:EMcorr}
\end{figure}

Figure~\ref{fig:EMcorr} shows the responses of the simple and $f_{\rm em}$ corrected energy sum reconstruction techniques on a sample pion initiated showers.  We can see that the $f_{\rm em}$ corrected energy sum largely restores the response linearity, as one expects from dual-readout.

%\subsection{CNN model}
\section{CNN Energy Reconstruction Methodology}
Complexity of the hadronic shower poses serious challenges to the analysis techniques targeting energy reconstruction based on the distribution of energy deposited throughout the calorimeter volume.

The nuclear interactions involved in hadronic showers open up many more possible signal generating processes compared to electromagnetic showers.

\begin{figure}[ht!]
\vspace{-1pc}
\centering
\includegraphics[width=0.88\textwidth]{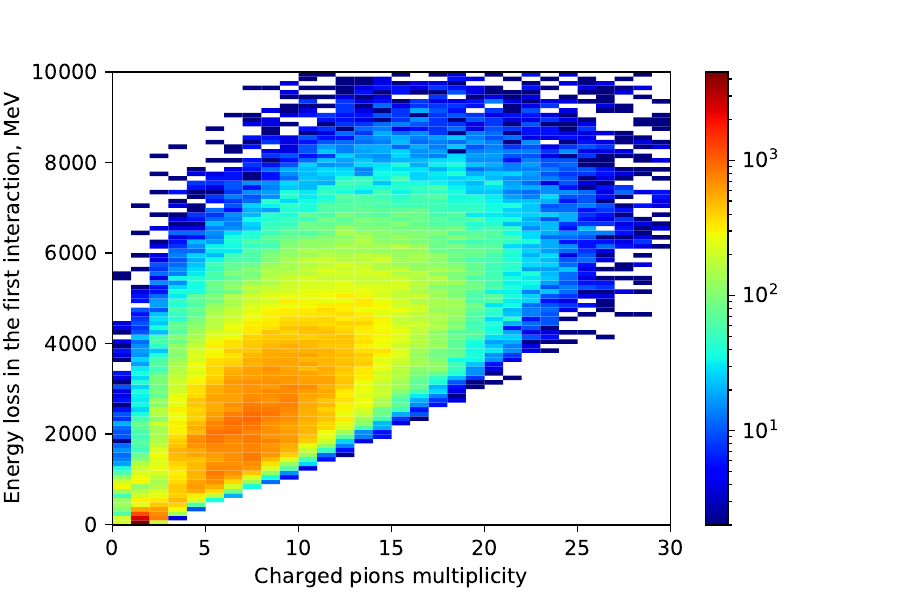}
\vspace{-0.5cm}
\caption{\small Energy loss at the first interaction is strongly correlated with the charged pion multiplicity in the hadronic showers as predicted by GEANT4.  Such relations between the invisible energy and specific features of the visible signal are utilized by neural networks for improved energy reconstruction. }
\label{fig:multip}
\end{figure}

Hadronic showers have complex correlation and dependencies amongst the observed signals because the various particle species emerging from a particular nuclear interaction subsequently deposit energy and participate in processes characteristic of the species.
For instance, the invisible energy in the hadronic component is strongly associated with the multiplicity of secondary charged hadrons as one can see in Figure~\ref{fig:multip}.
The distribution shown in Figure~\ref{fig:multip} does not appear as a simple correlation between the two variables; the distribution appears to have multiple modes and correlated strata which indicates we need some other variable or variables to fully explain this distribution.
This is only one example of one pair of shower attributes.

%\begin{wrapfigure}{r}{0.48\textwidth}
%\end{wrapfigure}

There are two main groups of secondaries with distinct topological properties. The first group encompasses secondaries from intra-nuclear scattering  - mainly light mesons ({\it i.e.} pions) which have a small deflection angle with respect to the direction of the initial hadron.
The second group consists of the product of nuclear de-excitation; its predominant contribution comes the from protons and neutrons emitted at significantly large angles.
High granularity calorimeters offer unique opportunities to detect topological signatures, such as tracks of the charged hadrons, and extract information to estimate, or compensate, for the undetected invisible energy. 

The common coincidence and overlap of the various topological signals, for instance EM showers often overlap the tracks from charged hadrons, makes it difficult to efficiently reconstruct the event with traditional techniques.
 
The array of cell responses of a highly granular calorimeter resembles a 3D image to which CNNs are well suited to extract high level properties of the objects found within.

\begin{figure}[ht]
\centering
\includegraphics[width=0.8\textwidth]{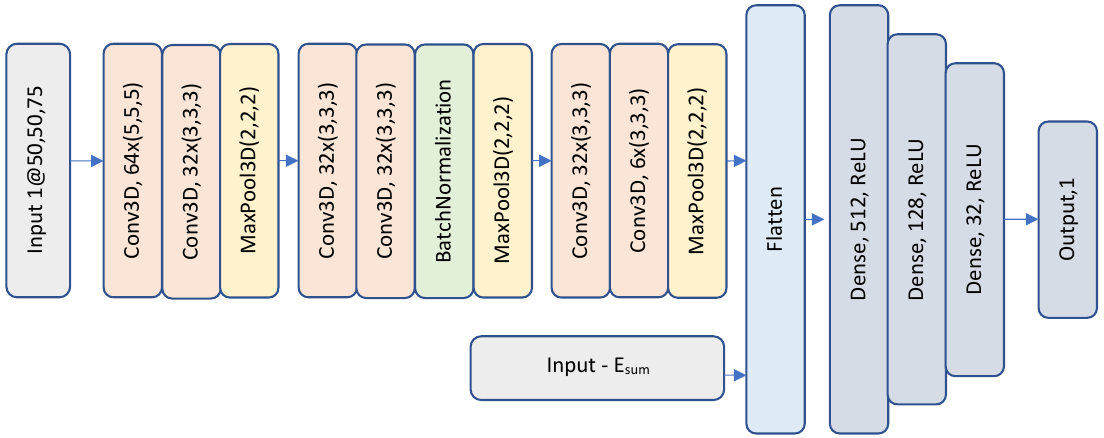}
\caption{The CNN architecture consists of several alternating 3D Convolutional and  MaxPooling layers in this work. Energy sum over the active volume is represented as an additional node in the flattened layer. Three Dense layers with dropout rate of 0.2 are used.}
\label{fig:CNN_Arch}
\end{figure}

We can treat the energy deposition in such a calorimeter as approximating a mono-chrome 3D image where the deposited energy represents the brightness level. The input image is a $50\times50\times75$ array of energy measurements.  The CNN in this analysis consists of 3D convolutional, max pooling, and flattened and dense layers as shown in Figure~\ref{fig:CNN_Arch}. The energy sum is represented as an additional input node in the flattened layer, which forces the CNN to work in energy correction regime and results in better overall response linearity.  

We optimize the mean square logarithmic error to estimate the model parameters to achieve more balanced contributions from low and high energies.  The Adam optimization algorithm~\cite{Adam2014} with learning rate 0.001 is used with early stopping based on the validation loss. Regularization in the training process is implemented with \texttt{BatchNormalization} layer~\cite{DBLP:journals/corr/IoffeS15} and \texttt{DropOut}~\cite{srivastava2014dropout} with a rate of 0.2 in-between the dense layers. 
In this configuration, the CNN energy reconstruction effectively works as an energy correction to the raw energy sum.  Correcting the raw energy sum allows the CNN to apply more universally to energy reconstruction of different particle types and multiplicities.  For example, the CNN derived energy correction in photon-initiated showers is negligible; the reconstruction performance is very similar to that of the simple energy sum since EM showers have few hadronic secondaries in the shower.  Samples of 600,000 simulated events are used for training. The validation is performed with independent sets with 50,000 events. The test samples have 300,000 events each. The optimal training is achieved within 15 epochs with the mentioned samples size. 

In the case of jets, the performance is expected to be similar to single-particle performance as the invisible energy in the particular hadronic interaction depends only on the initiating hadron and the media properties.

Studies based on Class Activation Mapping~\cite{zeiler2013visualizing} revealed that energy correction is mostly derived by the CNN in regions near individual hadron interactions with traces from charged hadrons and just outside regions of substantial EM energy depositions. These regions contain representative information on the multiplicity and production angle of the secondary charged hadrons.  These studies give us insight into how the CNN infers the undetected, invisible energy from the available visible signals.

%\newpage 
\section{CNN Performance with Single Hadron and Jets}
\label{sec:perf}

\begin{figure}[ht]
\centering
\includegraphics[width=0.455\textwidth]{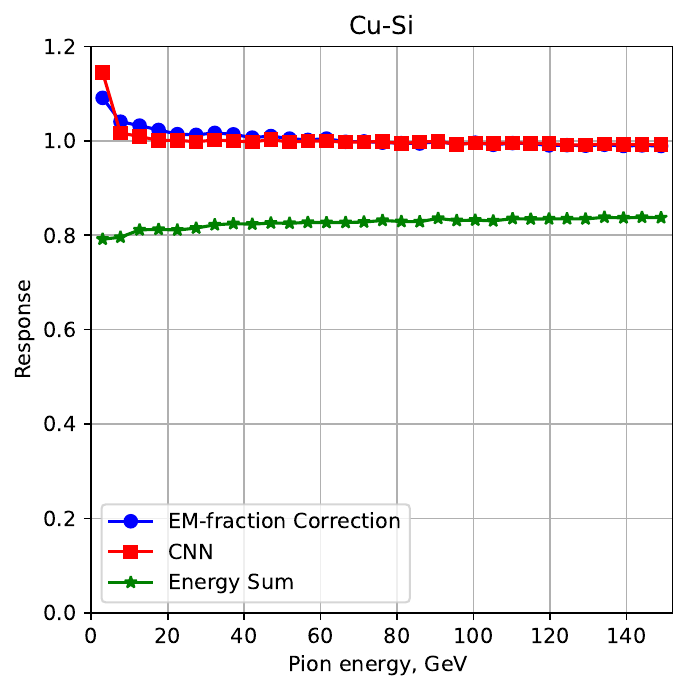}
\includegraphics[width=0.47\textwidth]{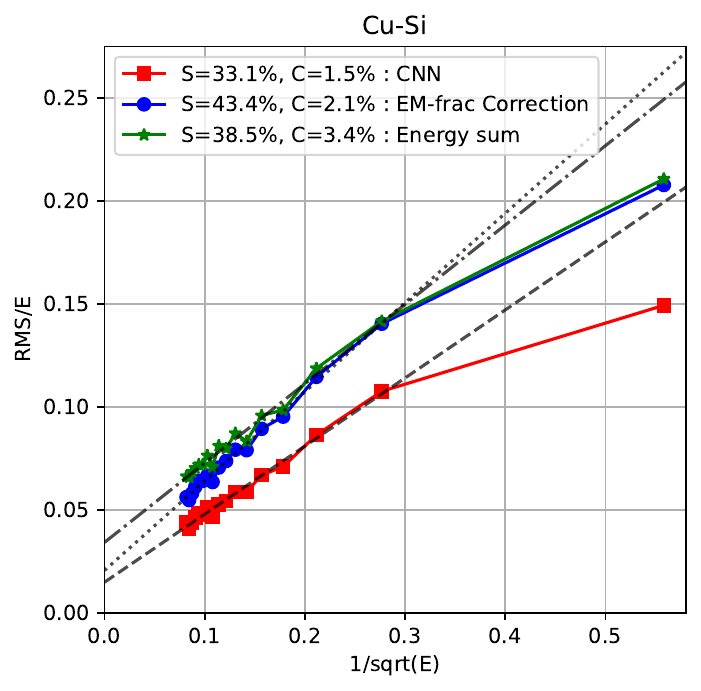}
\caption{The calorimeter response (left) and energy resolution (right) for charged pions are shown. The simple sum over all channels (green), with the $f_{\rm em}$ correction (red) and the CNN regression (blue) show respective energy measurement performance.  The $f_{\rm em}$ correction effectively employs the traditional dual-readout approach~\cite{AKCHURIN2005537}. The energy resolution parameters representing stochastic (S) and constant (C) effects estimated by a linear fit are included in the legend. }
\label{fig:CNN_Cu_pions}
\end{figure}

The 3D CNN is trained on a GEANT4-simulated data set with 0.5 to 150 GeV charged pions. The energy reconstruction performance is then tested on an independent sample in the same energy range.  
Figure~\ref{fig:CNN_Cu_pions} shows the reconstruction performance of the CNN compared against the simple energy sum and dual-readout approach.  We have included parametric energy resolution fits of the form $\frac{a}{\sqrt{E}} + b$ where $a$ and $b$ represent the stochastic and constant terms to the comparison plots.
One can see that the CNN outperforms both alternative reconstruction methodologies.
Here, the $f_{\rm em}$ correction method represents a dual-readout approach where $f_{\rm em}$ is computed from the energy deposited by the electrons in the shower.

\begin{figure}[ht]
\centering
\includegraphics[width=0.463\textwidth]{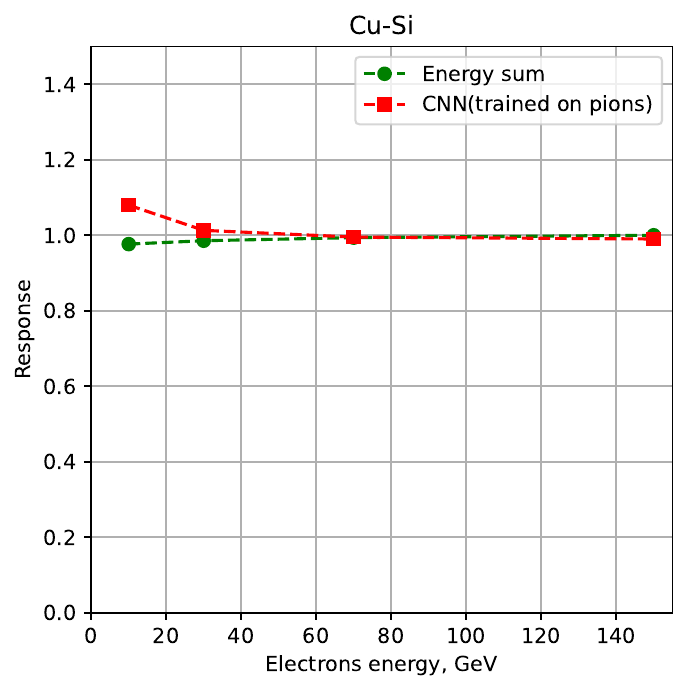}
\includegraphics[width=0.47\textwidth]{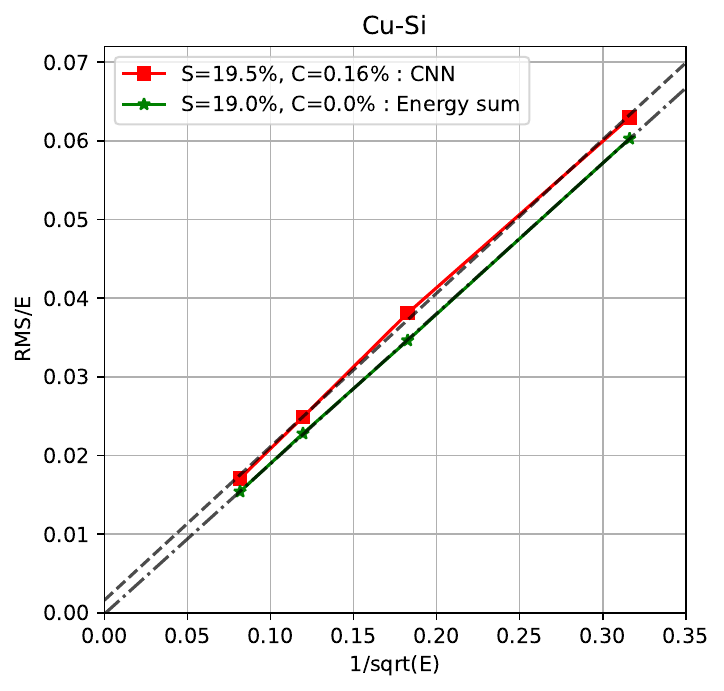}
\caption{The calorimeter response (left) and energy resolution (right) for electrons: the sum over all channels (green) and the CNN regression (red) result in similar performance
for a CNN that has been trained on pions alone.}
\label{fig:CNN_Cu_el}
\end{figure}

The CNN performance on electrons is illustrated in Figure~\ref{fig:CNN_Cu_el}; again the figure illustrates the performance of the simple sum for comparison. We see the CNN performance closely resembles, with minimal degradation, the performance of the simple energy sum. 
By the design of our CNN architecture we anticipated the similar performance because the CNN operates as an energy correction.

Thus, in the case of showers initiated by photons or electrons, the CNN does not find traces from charged hadrons and the correction to the raw calorimeter response is negligible. As a result, a CNN trained with pion data sets can reconstruct electron/photon energy without introducing a strong, undesired bias.
We did not study the performance of a CNN trained on a population comprised of pions and electrons.

\begin{figure}[ht]
\centering
\includegraphics[width=0.455\textwidth]{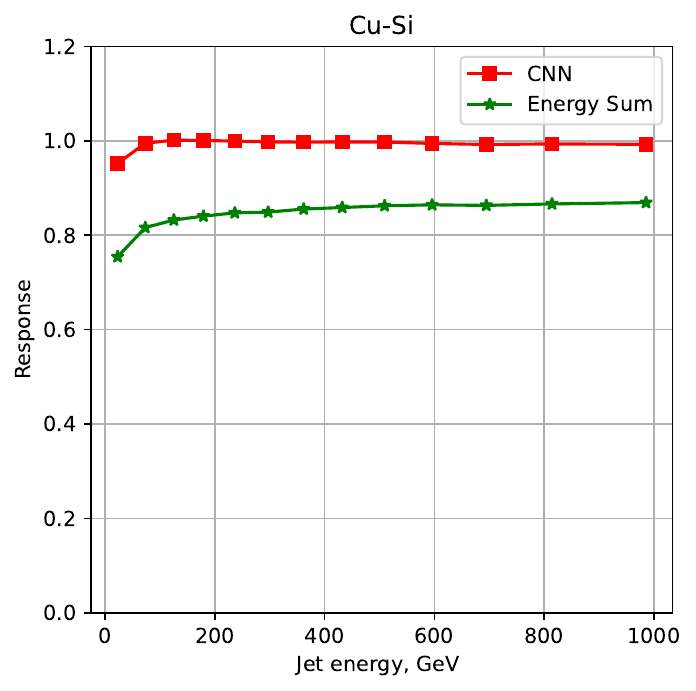}
\includegraphics[width=0.47\textwidth]{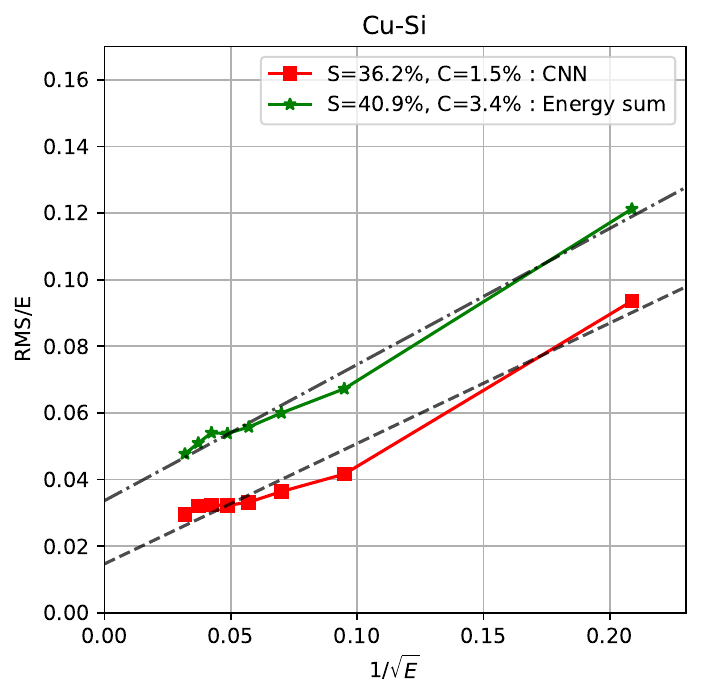}
\caption{The response (left) and energy resolution (right) for jets: the sum over all channels (green) and CNN regression (red).}
\label{fig:CNN_Cu_jets}
\end{figure}

We further examined the CNN reconstruction performance on jets by using PYTHIA8~\cite{Sjstrand:191756} to simulate $u$-quark jets with energy from 20 GeV to 1 TeV. The response linearity and energy resolution is shown in Figure~\ref{fig:CNN_Cu_jets}. The energy scale is preserved without the need for additional corrections. The energy resolution is also significantly improved when compared to the more traditional reconstruction techniques. 

\section{Reconstruction of $f_{\rm em}$ with CNN}
\label{sec:CNNfem}

The large fluctuations in $f_{\rm em}$ in non-compensating calorimeters is the leading source for performance degradation in energy reconstruction. Dual-readout calorimeters are designed to infer $f_{\rm em}$ on an event-by-event basis by using signals from scintillation and Cherenkov light. We have tested an alternative approach for $f_{\rm em}$ reconstruction in a single-readout calorimeter using a CNN that leverages topological information in the shower development. The CNN is trained on simulated charged pions 0.5-150 GeV to reconstruct $f_{\rm em}$ and $f_{\rm had}$.

\begin{figure}[ht]
\centering
\includegraphics[width=0.47\textwidth]{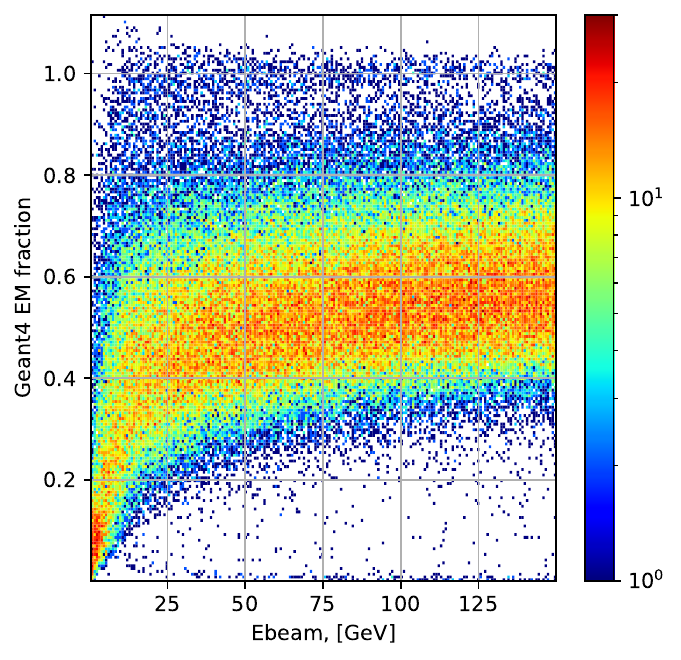}
\includegraphics[width=0.47\textwidth]{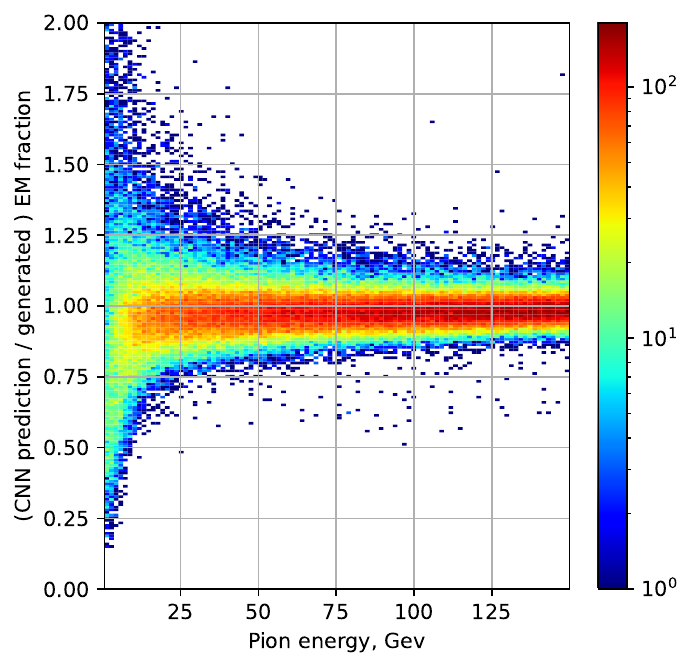}
\caption{The electromagnetic fraction in the energy deposit by pions (left) and the predicted EM fraction by the CNN normalized to the true value (right).}
\label{fig:CNN_Cu_pions_EMfrac}
\end{figure}

Figure~\ref{fig:CNN_Cu_pions_EMfrac} shows the simulated $f_{\rm em}$ (left) and the ratio of the reconstructed to simulated $f_{\rm em}$ (right) over the range of particle energies.  One can deduce the viability of reconstructing $f_{\rm em}$ in a single readout high granularity calorimeter from the result illustrated in this figure.

\section{Beyond the $f_{\rm em}$ correction}
\label{sec:USi}
We also studied the energy reconstruction using a compensating ($e/h = 1$) calorimeter where the $f_{\rm em}$ fluctuations are no longer the leading cause for degraded performance.

\begin{figure}[ht]
\centering
\includegraphics[width=0.455\textwidth]{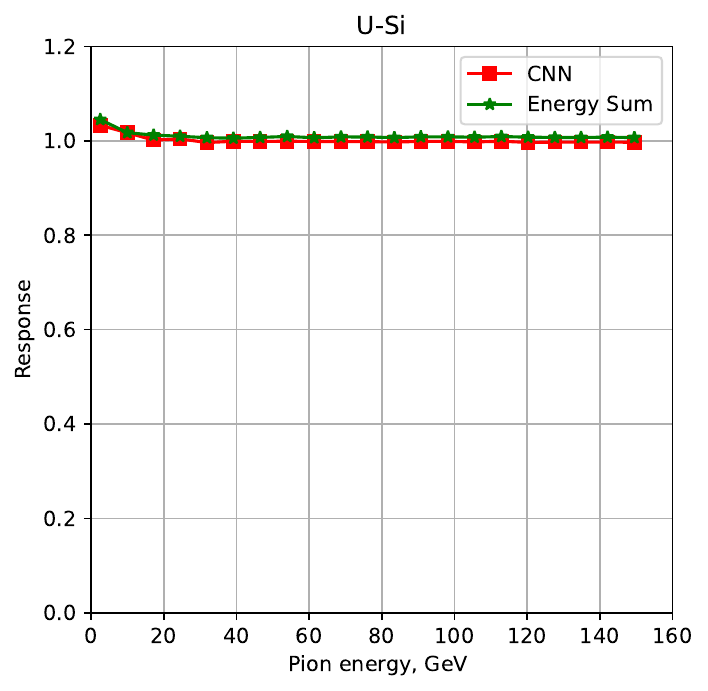}
\includegraphics[width=0.47\textwidth]{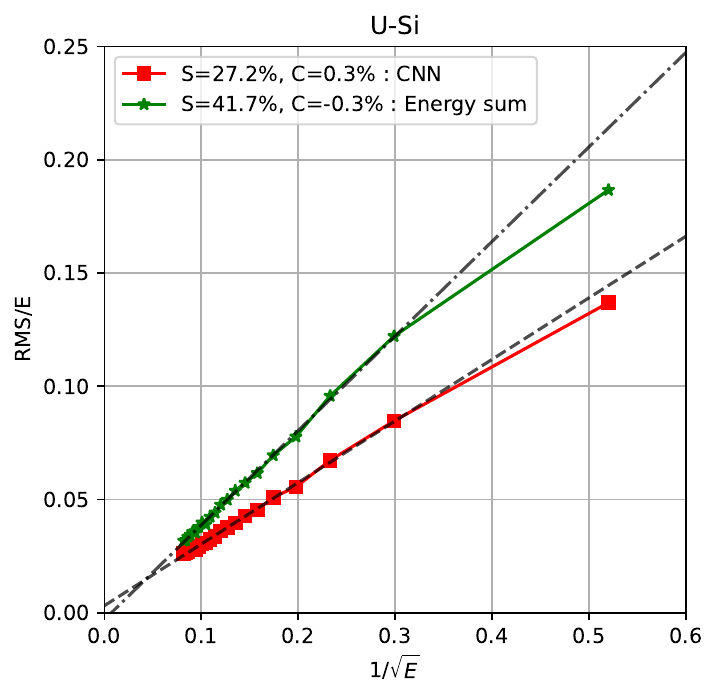}
\caption{The response (left) and energy resolution (right) for pions: the sum over all channels (green) CNN regression (red).}
\label{fig:CNN_U_pions}
\end{figure}

The simulated U-Si calorimeter shows linear response to pions using simple sum for energy reconstruction.
The improved reconstruction performance of the CNN over the simple energy sum as shown in Figure~\ref{fig:CNN_U_pions} indicates that the CNN exploits the relationship between the invisible energy and the visible signal in the shower, {\it i.e.} the multiplicity and production angle of the secondaries from the hadron interactions.

\newpage 
\section{Using Timing Information with Graph Neural Networks (GNN)}
\label{sec:timing}

Graph Neural Networks (GNNs) can cope with irregular detector geometries which make them potential candidates for use in a full-size collider based experiment.
Further, GNNs allow for the assignment of an arbitrary number of properties to a cell, thereby allowing us to easily incorporate multi-readout and timing information.
%Implementation of the timing information as colors in the CNN did not produce promising results. 
We chose Dynamic Graph CNN~\cite{Wang3326362} to investigate the impact of precise timing for energy reconstruction.The configuration includes four EdgeConv blocks with two-layer MLP ([64,64], [96,96], [128,128], [256,256]) and k-NN parameter set to 10. We use the same training parameters and samples as described for the CNN training. 
We studied the GNN performance on the same Cu/Si setup described above.

\begin{figure}[ht]
\centering
\includegraphics[width=0.32\textwidth]{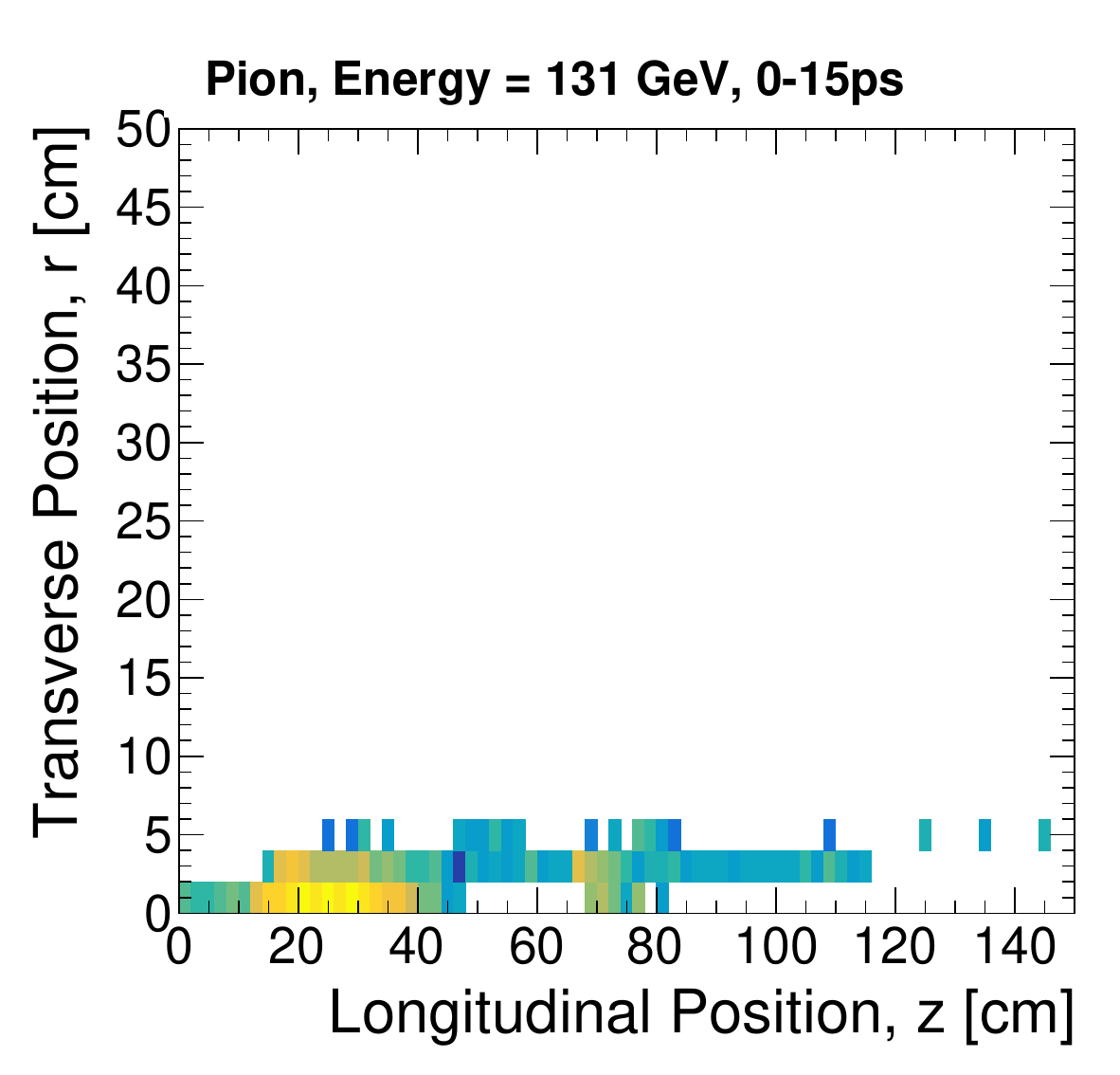}
\includegraphics[width=0.32\textwidth]{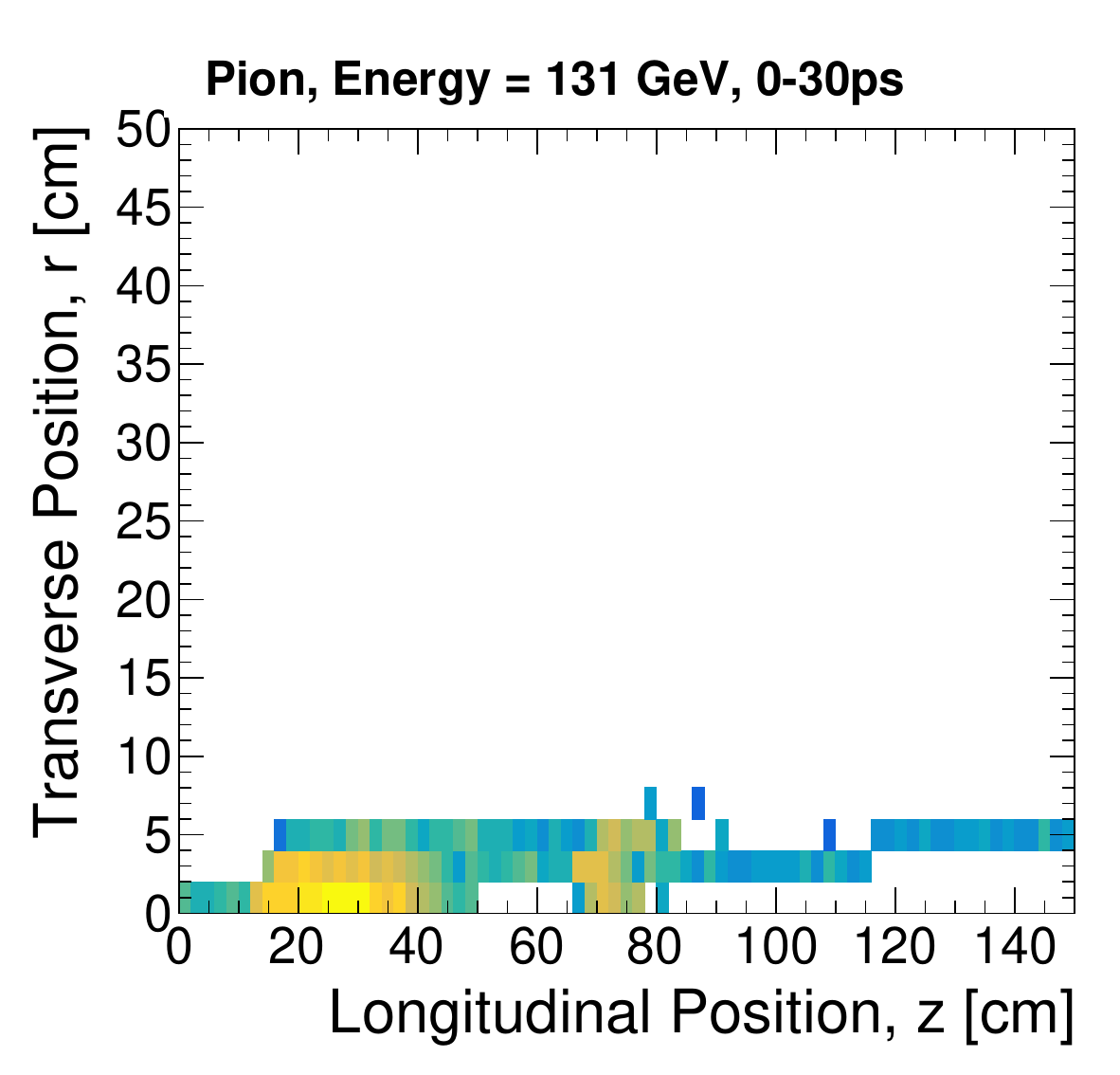}
\includegraphics[width=0.32\textwidth]{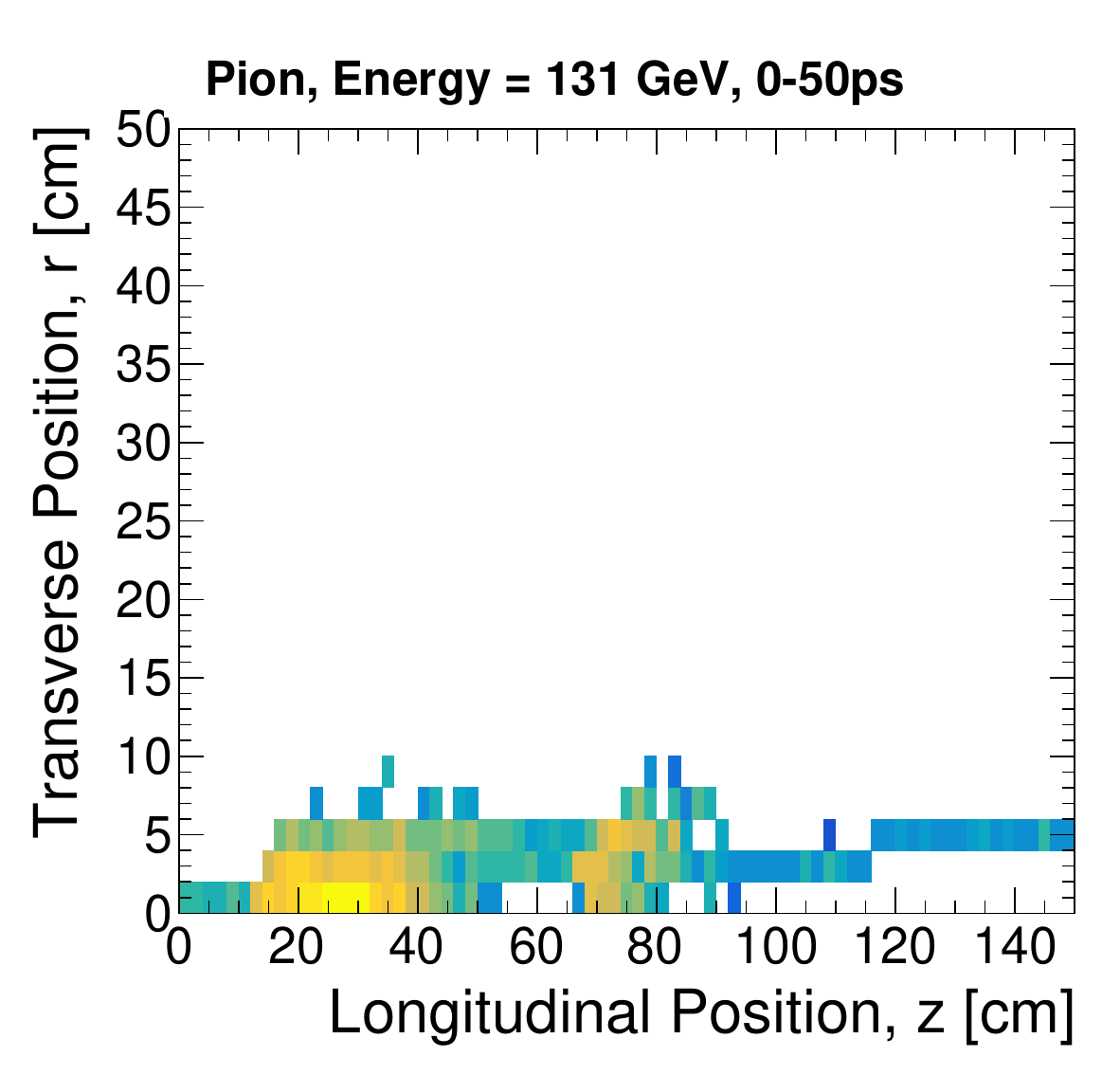}\\
\includegraphics[width=0.32\textwidth]{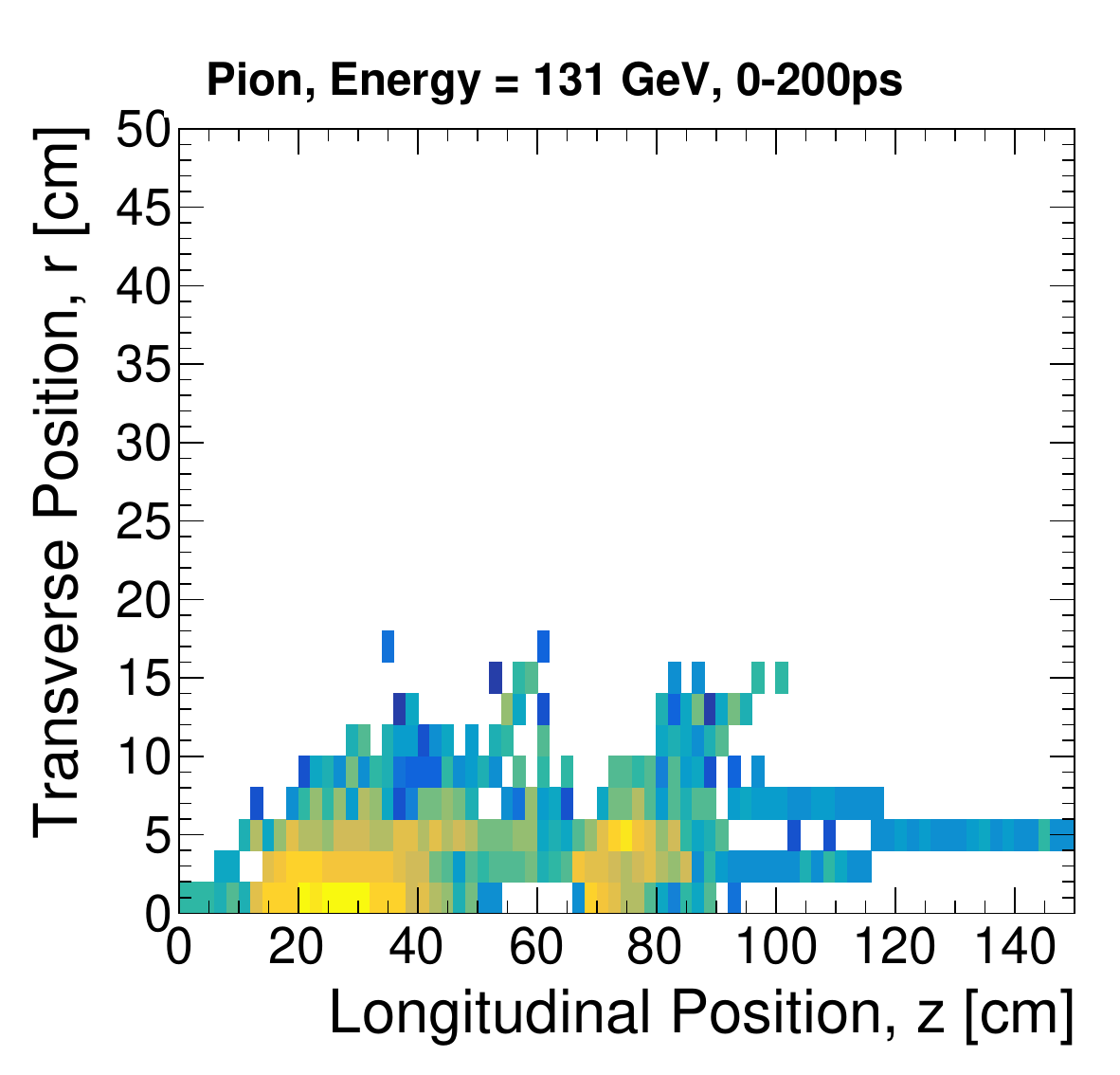}
\includegraphics[width=0.32\textwidth]{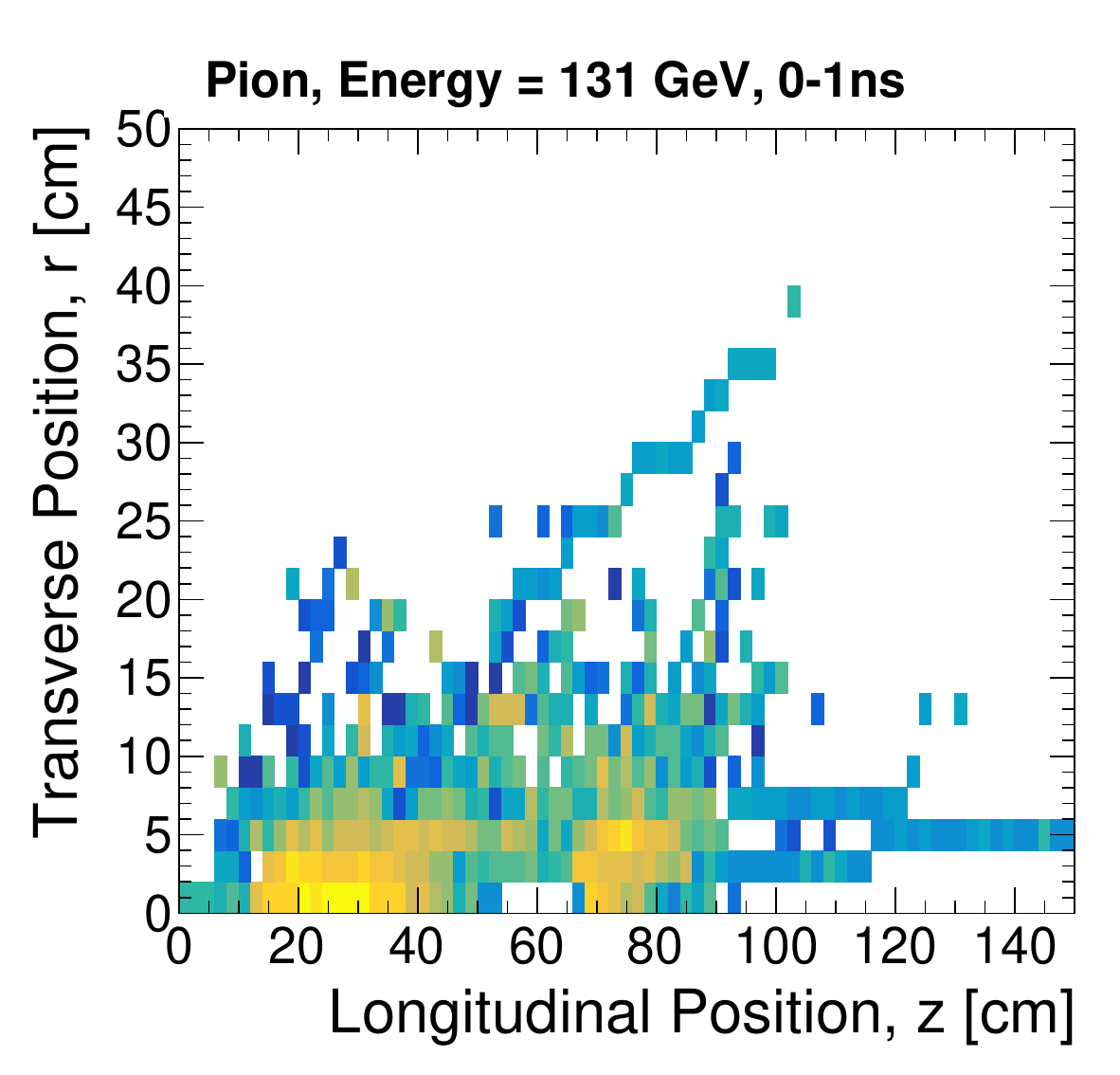}
\includegraphics[width=0.32\textwidth]{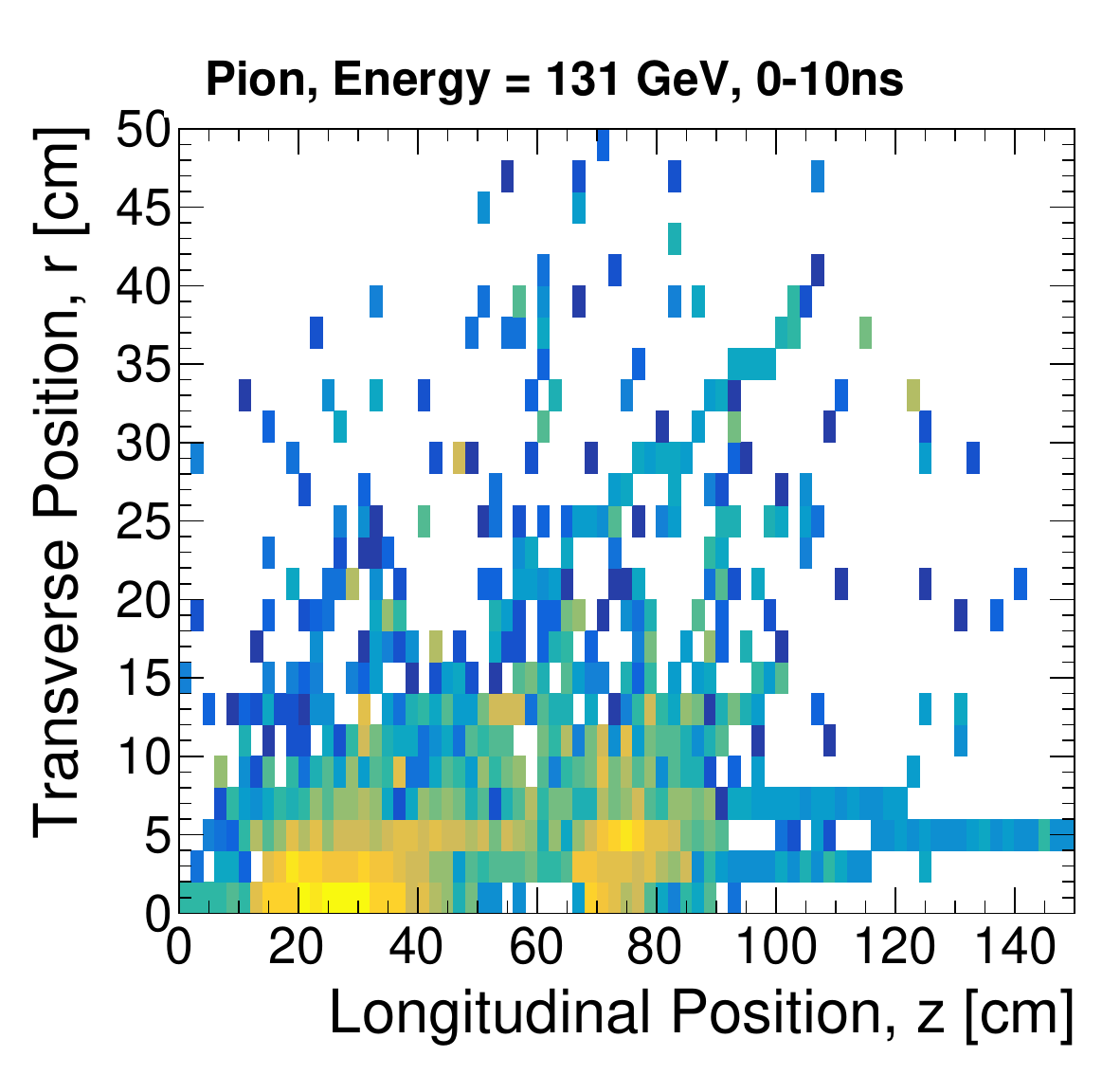}

\caption{The energy deposits due to a single 131 GeV charged pion are shown in $r-z$ coordinates where the colors indicate deposited energy.  As indicated on top of each plot, the integration times gradually increase from 0-15 ps (top left) to 0-10 ns (bottom right). Time is `local', in other words, it is corrected for the travel time, $t = t_{\rm G4} - z/c$,  along $z$-axis for all particles. }
\label{fig:ShowerTimePi}
\end{figure}

\begin{figure}[ht]
\centering
\includegraphics[width=0.32\textwidth]{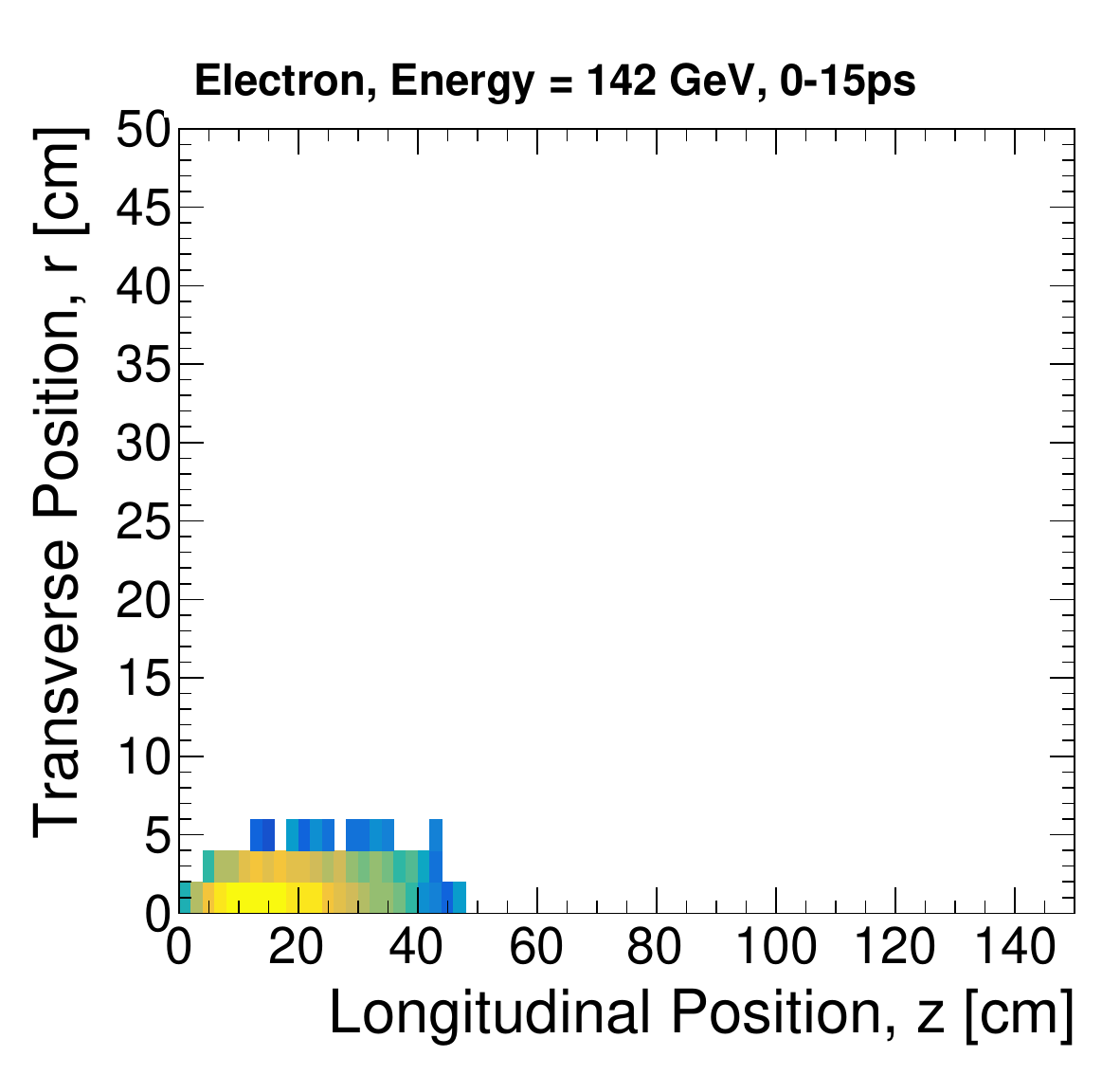}
\includegraphics[width=0.32\textwidth]{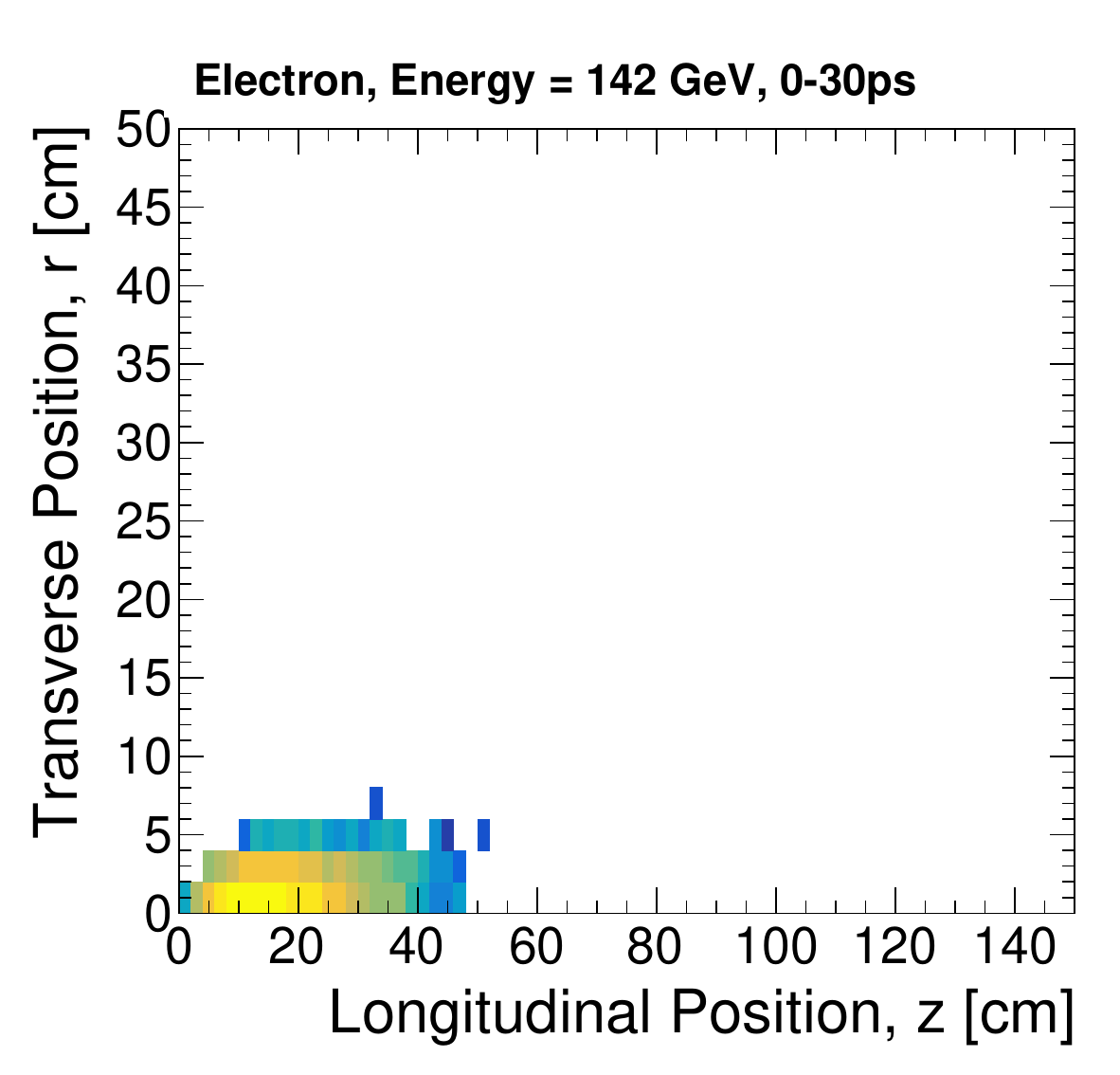}
\includegraphics[width=0.32\textwidth]{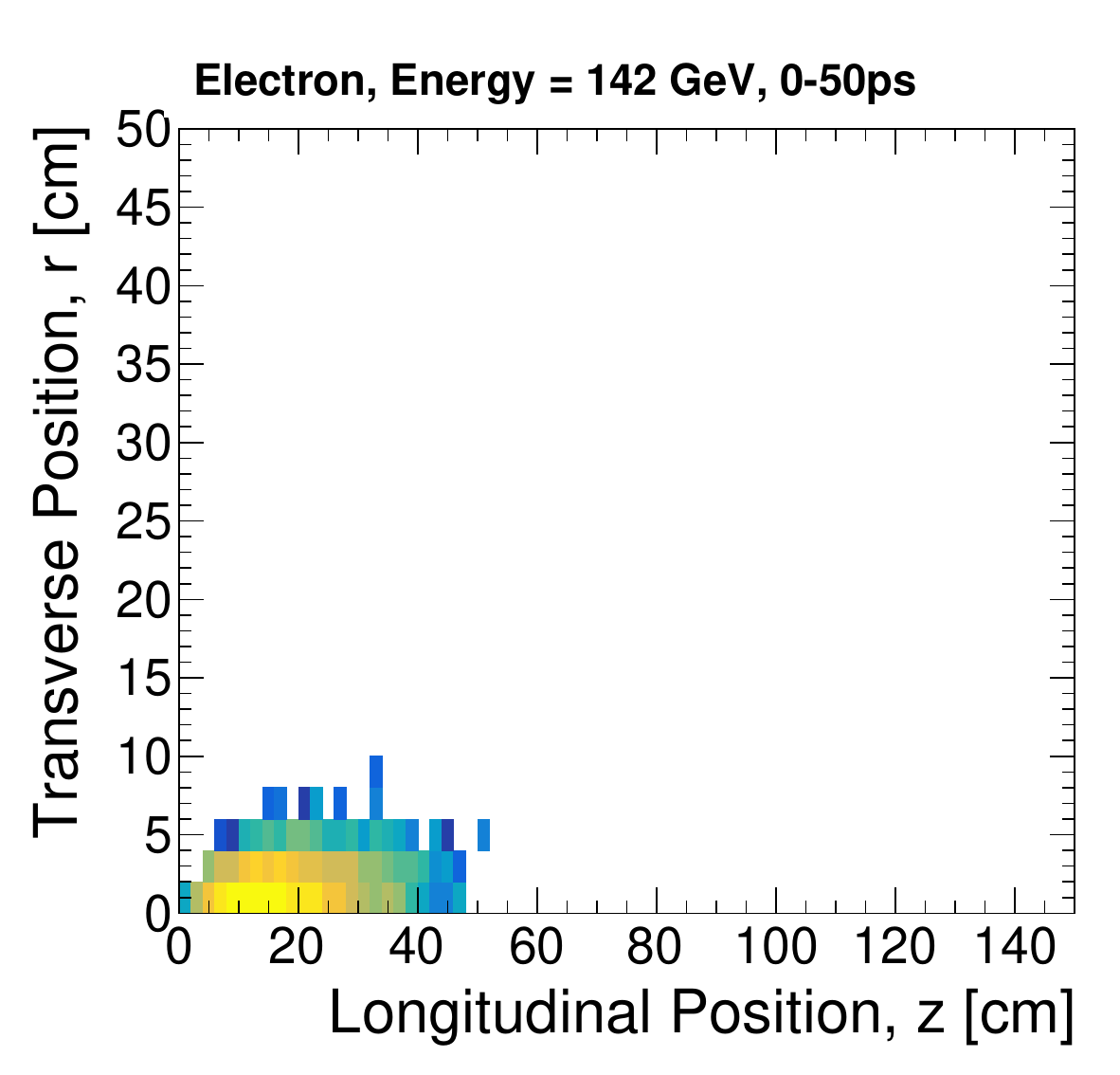}\\
\includegraphics[width=0.32\textwidth]{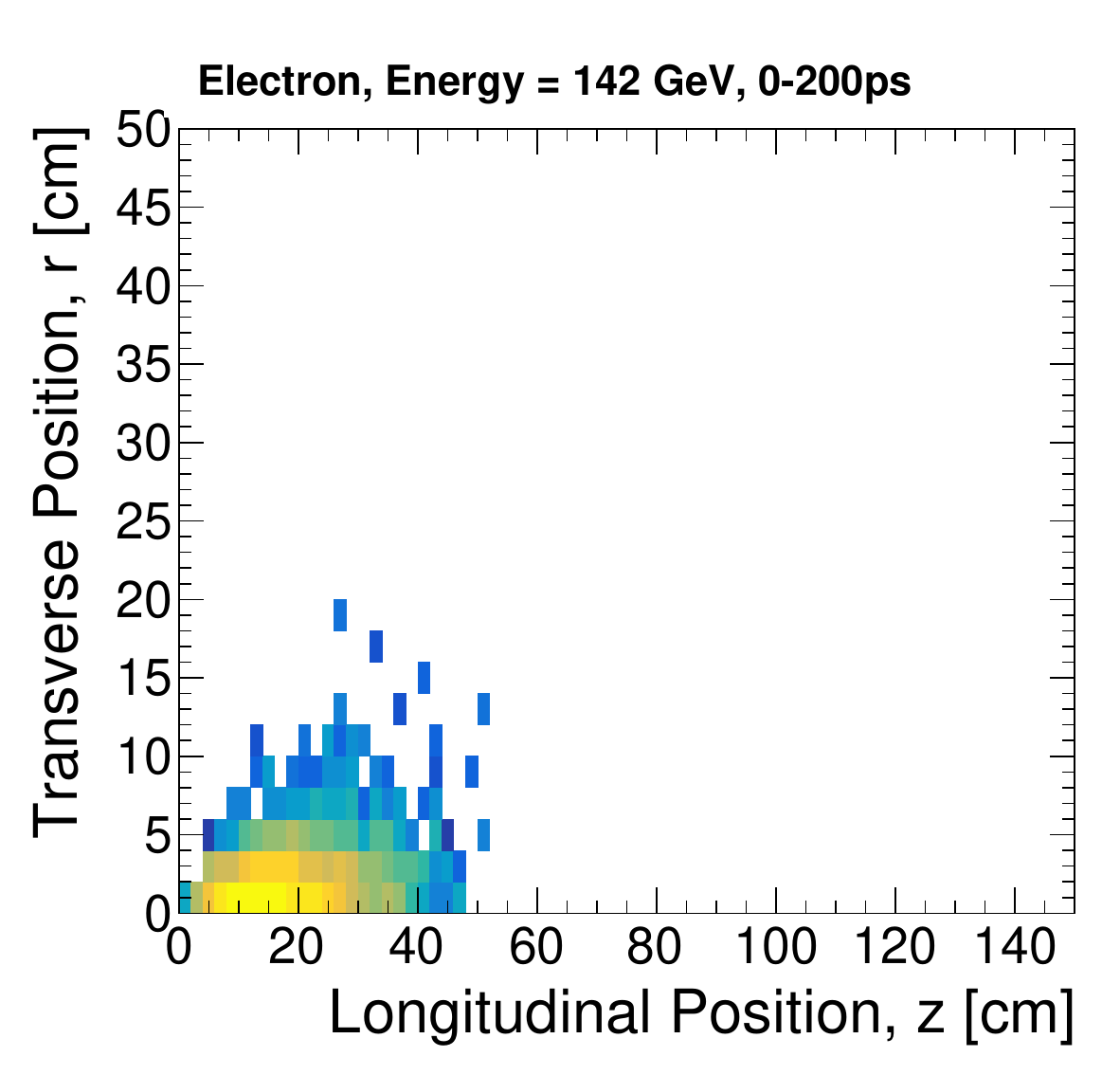}
\includegraphics[width=0.32\textwidth]{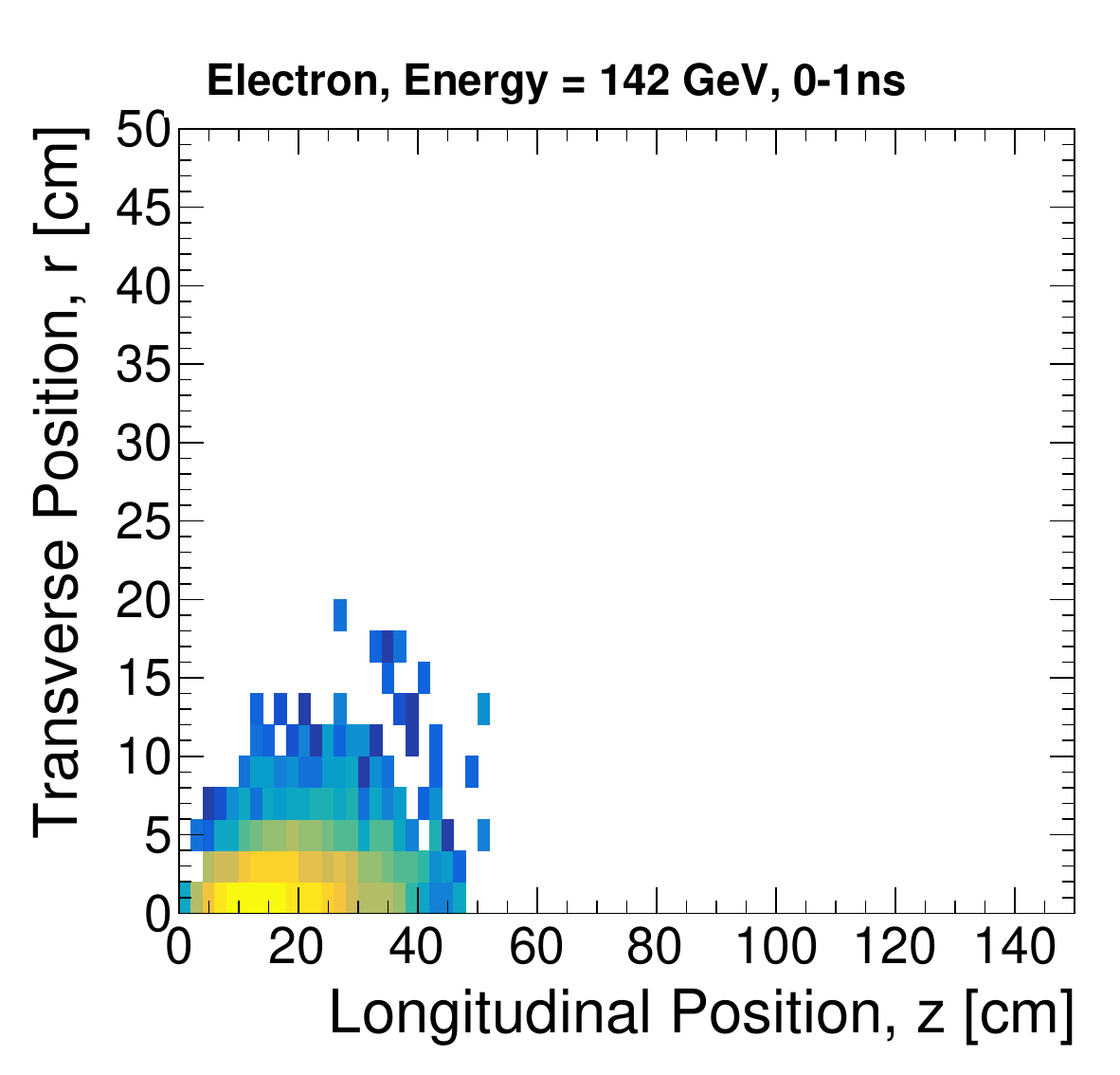}
\includegraphics[width=0.32\textwidth]{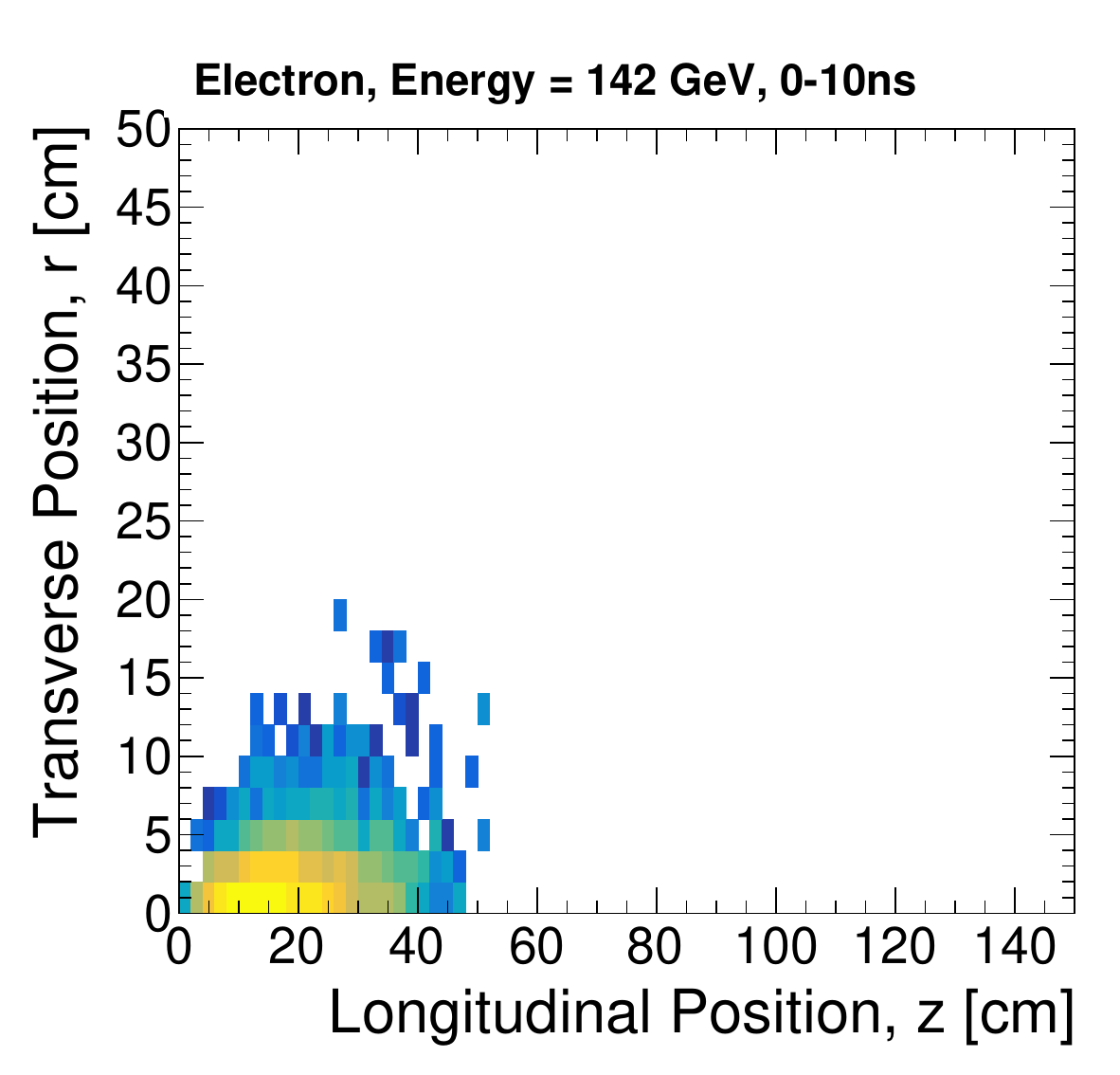}
\caption{The energy deposits due to a single 142 GeV electron are shown in $r-z$ coordinates where the colors indicate deposited energy.  As indicated on top of each plot, the integration times gradually increase from 0-15 ps (top left) to 0-10 ns (bottom right). Time is `local', in other words, it is corrected for the travel time, $t = t_{\rm G4} - z/c$,  along $z$-axis for all particles.  }
\label{fig:ShowerTimeEl}
\end{figure}

%The time of energy depositions in the individual cells is extracted from the Geant4 simulations, where the travel time in the longitudinal direction ($z$) subtracted assuming speed of light in vacuum ($t = t_{G4} - z/c$).  In any particular cell there could be multiple deposits from different particles in the shower, which defines the time-profile of the signal. Then we calculate energy deposition in different integration time intervals for each cell to study the importance of the shower timing for the energy reconstruction. The energy depositions in several different time intervals are used in the features set of the GNN. The choice of time intervals is described later in the text. 
We feed into the GNN a series of cell energy readouts having increasing integration times.  In this way, the series of cell energies represent cumulative effects of time rather than distinct time bins.  The energy is accumulated up to a final 10 ns integration interval, and we leave the number of time slices as a variable to study the impact of timing precision.  We record the time of any simulated energy deposition as $t = t_{\rm G4} - z/c$ where $t_{\rm G4}$ is the time when the energy is deposited as reported by GEANT4 and $z/c$ is the travel time of light in vacuum to cover the longitudinal depth.  Single pion and electron event energy distributions in space for various integration times\footnote{We denote the duration of the window of time a signal may be collected and observed as the integration time.} are shown respectively in Figures~\ref{fig:ShowerTimePi} and~\ref{fig:ShowerTimeEl}.  

Each subfigure gives a representation of the spatial distribution of energy at some integration time; the $y$-axis represents the radial distance from the shower axis and the $x$-axis represents the longitudinal depth of the shower.  We chose this form of visualization to draw attention to the development of the radial extent of showers at different time scales.

Unlike hadronic showers, the electron initiated energy depositions take place promptly without much structure; low energy photons form the energy deposits found deep or transversely far from the shower axis in the calorimeter.
Roughly speaking, by going from long to short integration times, we effectively transversely segment this calorimeter.
This segmentation supplies important information since the time scales of some hadronic processes are much longer than the EM processes.

\begin{figure}[ht!]
\vspace{-1pc}
\centering
\includegraphics[width=0.88\textwidth]{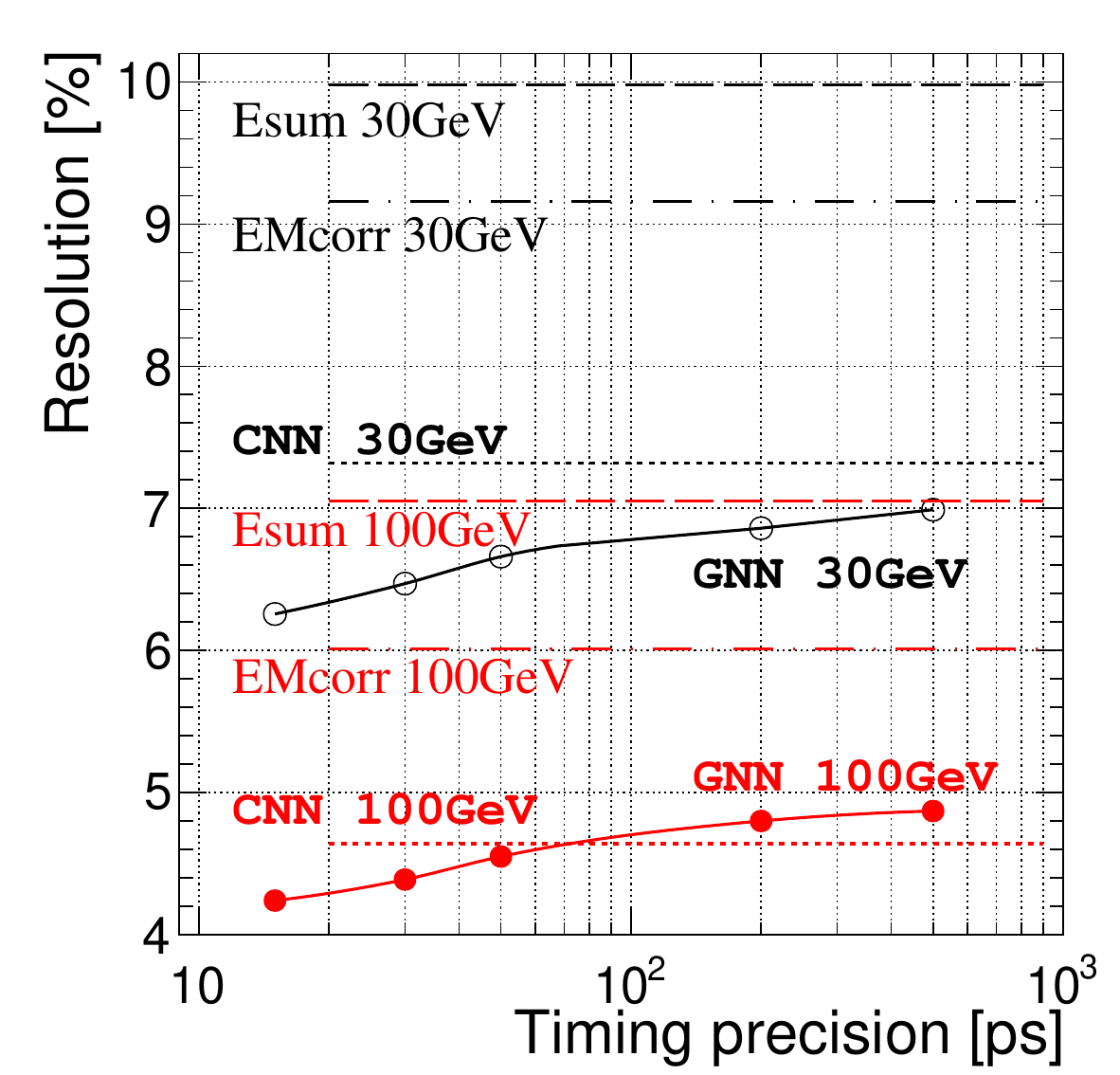}
\vspace{-0.3cm}
\caption{\small The energy resolution ($\sigma/E$) for 30 GeV (black) and 100 GeV (red) pions. Simple energy sum (Esum), $f_{\rm em}$ corrected energy sum (EMcorr), CNN and GNN reconstruction techniques. The horizontal axis indicates the assumed timing precision for the GNN technique.  The energy resolutions obtained from different reconstruction techniques are also shown for comparison. }
\label{fig:GNNperf}
\end{figure}

Figure~\ref{fig:GNNperf} illustrates the effect of increasingly better timing measurement on energy resolution. Each data point (solid and open circles) includes several time slices.  While the signal is integrated for 10 ns for all data points, the precision with which timing intervals are known is plotted on the horizontal axis.  For example, the timing precision of 0.5 ns includes time intervals (0-0.5 ns), (0, 4 ns), (0, 1 ns) and (0, 0.5 ns) and plotted at 0.5 ns.  As shorter time slices are included, the energy resolution improved somewhat for the two cases (30 and 100 GeV pions) we analyzed.  We posit that information inherent in short time slices, possibly due to low energy protons, contribute additional information to the network.  Essentially, protons in shorter time scales seem to stand in for neutrons in longer time scales for enhanced energy measurement.

The GNN selects the local region of interest by a fixed number of cells with non-zero energy ($k$ Nearest Neighbours). The size of the region varies with the density of the active cells, which is also correlated with the energy of the incident particle. In contrast the CNN, it uses a fixed size for the local region of interest defined by the filter size. This leads to the observed difference in the performance between CNN and GNN as a function of the particle energy and is subject of further  development and tuning of these algorithms. In this study, we mostly focused on exploring possible improvement in energy reconstruction by the addition of shower timing information. 

%\begin{wrapfigure}{l}{0.48\textwidth}

%\vspace{-4pc}
%\end{wrapfigure}

\section{Remarks}
\label{sec:conclusion}

In conclusion, we have observed enticing performance results from our simulation study of advanced energy reconstruction methodologies applied to a high granularity calorimeter module.
The CNN response to pions appears quite linear at high energies, it tends towards unity faster than the simpler method based on $f_{\rm em}$, and it also significantly outperforms both the energy sum and $f_{\rm em}$ methods in terms of energy resolution.
By regressing, {\it e.g.} estimating, $f_{\rm em}$ and $f_{\rm had}$ with a CNN model we add further support to our deduction that the evidence of these fluctuations exists within the spatial distribution of deposited energy.  Similarly, the improved energy resolution noted in the study of a compensating ($e/h = 1$), U-Si module suggests the CNN's sensitivity to $f_{\rm had}$.
Altogether, this evidence suggests that the CNN/GNN methodology applied in high granularity calorimeters surpasses the hadronic energy resolution attainable with a comparable dual-readout device.
Finally, we studied the prospects of a GNN applied to our simulated calorimeter module with the addition of precision timing information.  We observed a time precision dependence of energy resolution - better timing, better resolution - roughly comparable to that of the CNN.  The GNN's energy resolution surpasses the CNN's below $\sim 100\,{\rm ps}$. 

% what it means
We have shown clear performance benefits by application of advance energy reconstruction techniques in a high granularity calorimeter.  We have also seen evidence which suggests that the neural network type reconstruction models take advantage of the spatial, and spatial-temporal, distribution of deposited energy to realize such gains.  Although the precise way in which the CNN accounts for $f_{\rm em}$ and $f_{\rm had}$ remains unknown, we have shown that the CNN's sensitivity by showing that it can estimate these fractions based on the energy distributions.
This observation carries important implications for multi-readout calorimetry since the capability to estimate $f_{\rm em}$ and $f_{\rm had}$ from a single readout device may be at hand.

% what may come
Future studies will quantify the performance gains sustained when the simulation includes electronic noise, intercalibration, and other issues which impact physical devices.  One may pessimistically anticipate the CNN performance to degrade as noise increases and input response becomes less homogeneous; however, we have seen such comfortable margin of improvement over the simple energy sum and dual readout analog that we can still expect some good degree of improvement.  

\section{Acknowledgements}
This work has been supported by the US Department of Energy, Office of Science (DE-SC0015592) and Texas Tech University, Office of the Vice President for Research and Innovation.
\newpage
\bibliography{TTUbibfile.bib}
\end{document}